\documentclass[11pt,a4paper]{article}
\usepackage[utf8]{inputenc}
\usepackage[english]{babel}
\usepackage{amssymb, amsmath, enumerate}
\usepackage{graphicx}
\usepackage{epstopdf}
\usepackage{subcaption}
\usepackage{authblk}
\usepackage{textcomp}

\newcommand{\qed}{\nobreak \ifvmode \relax \else
      \ifdim\lastskip<1.5em \hskip-\lastskip
      \hskip1.5em plus0em minus0.5em \fi \nobreak
      \vrule height0.75em width0.5em depth0.25em\fi}

\begin{document}
\title{Parallel computing as a congestion game}
\author[1]{O.A.~Malafeyev\thanks{o.malafeev@spbu.ru}}
\author{S.A.~Nemnyugin\thanks{s.nemnyugin@spbu.ru}}
\affil[1]{Saint-Petersburg State University,  Russia}
\date{}
\maketitle
\begin{abstract}
Game-theoretical approach to the analysis of parallel algorithms is proposed. The approach is based on presentation of the parallel computing as a congestion game. In the game processes compete for resources such as core of a central processing unit and a communication subsystem. There are players, resources and payoffs (time delays) of players which depend on resources usage. Comparative analysis of various optimality principles in the proposed model may be performed.
\end{abstract}



\section{Introduction}
Development of optimal control methods is of great importance for different problems, see for example \cite{MalNem001}, \cite{MalNem002} and \cite{MalNem003}. One of actual such problems is high-performance computing which is complex problem including both hardware (see for example~\cite{Zakharov}) and software\cite{Boulychev}. Execution of parallel applications may be considered as complex interaction of few components. First one is set of parallel programs queued on high-performance computing cluster. Each parallel application is constituted of processes with communications between them. Second component is parallel computing system which consists of computing nodes and communication subsystem. A computing node provides central processor and memory to executing programs. Main component of the software environment is operating system with scheduler which orchestrates set of processes placed on the node. Due to limited hardware resources parallel computing may be considered as a competition of a set of processes for resources. Such competition should not decrease overall efficiency of the whole system so search of optimum and analysis of various optimality principles is of significant importance for the control of parallel computations.

We propose game-theoretical approach to the analysis of parallel algorithms.  Shortly it may be formulated as follows. There are players (processes), resources and payoffs (time delays) of players which depend on how much of a resource is used by processes placed on the computing node. It is the case of congestion or, more general, potential game\cite{iarieli}. Results of congestion games theory was used in consideration of networking problems \cite{tekin}. Congestion game theory results may be used to derive some conclusions about efficient structure of parallel algorithms and their dispatching on parallel computing systems. 

\section{Model of parallel computing}
Algorithm of a parallel application process may be presented by directed acyclic graph (DAG) \cite{Voevodin}. In the article we will refer to the DAG as information graph of an algorithm. Nodes of an information graph are macrooperations. The macrooperation transforms input data into output and send result to next macrooperation or a final result to the user. Links present data transactions between macrooperations. We consider case when only one source $s$ (data input) and only one sink $t$ (data output) are associated with any information graph: $PG = (V, L), s, t \in V$. Here $V$ is set of nodes of the graph and $L$ is set of directed links.

\begin{figure}[h!]
		\center{\includegraphics[scale=0.37, bb=385 5 83 300]{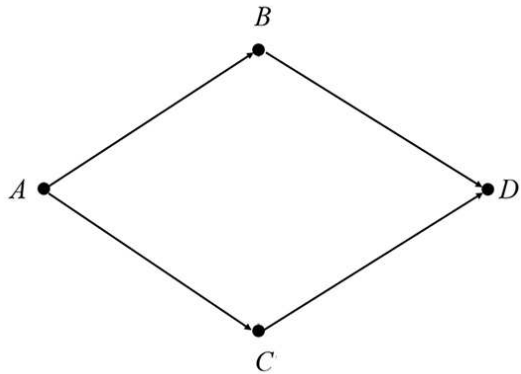}}
		%	\center{\includegraphics[scale=0.27]{fig001.eps}}
		%	\vspace*{1.0cm}
		\caption{Information graph of sequential or simple parallel algorithm}
		\label{ris:image1}
	\end{figure}

At first let us consider case of sequential program. Its graph is given, for example, by upper half of the graph in  Fig.~\ref{ris:image1}. Here $s \equiv A$ is source which receives input data and $t \equiv D$ is sink which performs output operations of the resulting data. In the graph two branches exist: $ABC$ and $ACD$. Both branches correspond to execution of the same sequential program on two different computing nodes. If the target computing system is symmetrical this two cases are equivalent. In the case of asymmterical system the branch is preferrable which corresponds to execution on the node with higher performance. 

One of the most important metrics of a parallel application is its total time of execution which may be defined as maximum of execution times over all processes. Now let us suppose that Fig.~\ref{ris:image1} is information graph of the algorithm which is almost ideally parallel and macrooperations $B$ and $C$ have equal computational complexities. Again let us consider the case of symmetrical target computing system. Here optimum in time of execution has place when data flow from $A$ is divided into two equal parts. It is Nash equilibrium because if one player, for example, corresponding to the branch $ACD$ tries to improve time of its own execution by decreasing amount of data then the second player ($ABC$) have to process greater amount of data. Overall performance in this case will be less then in equilibrium. In general, the Nash equilibrium is a set of strategies in a non-cooperative game when each player knows the equilibrium strategies of the other players, and no player can't get additional benefit by changing only its own strategy. In general Nash equilibrium doesn't guarantee best performance.
\vspace{1cm}

\begin{figure}[h]
	\center{\includegraphics[scale=0.37, bb=385 5 83 400]{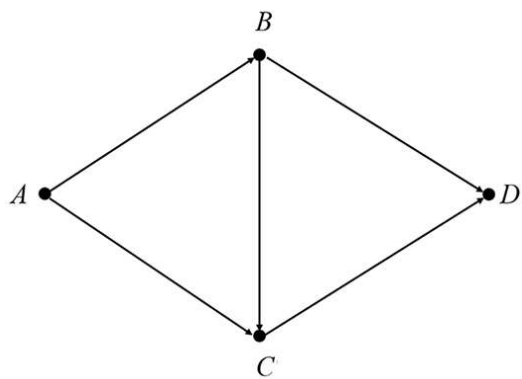}}
	%%%	\center{\includegraphics[scale=0.27]{fig002}}
	%	\vspace*{1.0cm}
	\caption{Information graph of parallel algorithm}
	\label{ris:image2}
\end{figure}

Next let us consider information graph  (Fig.~\ref{ris:image2}). It is the classical simplest example where the Braess paradox may take place \cite{MalKol}. Now three paths of execution there are exist: $ABD$, $ACD$ and $ABCD$. In this case $BD$ and $BCD$ may be placed onto different computing nodes.

In the model we propose in the article influence of other processes placed on computing nodes have to be taken into account. Due to competition for limited resources the more is the load of the node the more is time of execution of the process. It is the case of a congestion game. Generally, software developer takes care for minimization of time of execution of every branch of a parallel algorithm. In game-theoretical language it may be interpreted as selfish behaviour of players. A congestion game is $CG=(G, (s,t), (f_e)_{e \in L})$, where $f_e$ is nonnegative, continuous and nondecreasing latency function and $G$ is graph of an algorithm. Latency function describes slowing down of process execution due to dependence of the time of computation on the data volume and possibly from other factors. In more general case in latency function may be included accounting of the competition of different processes for limited resources. 
Total latency function is alsow defined as follows: $LF_{tot}(x)=\sum_{e\in \Phi}f_e(x)$, where $\Phi$ is a set of links in the path of execution for the given parallel algorithm. $LF_{tot}(x)$ may be taken as a criteria of optimization.

In the model described existence and uniqueness of Nash equilibrium should be studied. The case with large amount of data or players may be approximated by the game with infinite number of players. Analogue of Nash equilibrium in the game with infinite number of players is Wardrope's equilibrium \cite{MalKol}. 
\section{Conclusion}
In the article game-theoretical approach to the analysis of efficiency of parallel algorithms is proposed. It is assumed that parallel computing may be considered as congestion game with finite or infinite number of players. Principles of optimality may be formulated for such game and different kinds of equilibrium should be studied. It is the topic of further study.


\begin{thebibliography}{}  % (do not forget {})
	
	\bibitem{MalNem001}
	Malafeyev~O.A., Neverova~E.G., Nemnyugin~S.A., Alferov G.V.  Multi-criteria model of laser radiation control // 2014 Tenth IVESC 
	and Second International Conference on Emission Electronics ICEE / Proceedings edited by N. V. Egorov. Piscataway: IEEE, 2014. P.33.
	
	\bibitem{MalNem002}
	Malafeyev~O.A., Nemnyugin~S.A., Alferov~G.V., Charged particles beam focusing with uncontrollable changing parameters // 2014 Tenth IVESC 
	and Second International Conference on Emission Electronics ICEE / Proceedings edited by N. V. Egorov. Piscataway: IEEE, 2014. P. 25.
	
	\bibitem{MalNem003}
	Malafeyev~O.A., Alferov~G.V., Andreyeva~M., Group strategy of robots in game-theoretic model of interception with incomplete information. Source of the Document
	2015 International Conference on Mechanics - Seventh Polyakhov's Reading. 2015.
	
	\bibitem{Zakharov}
	Stepchenkov Y.A., Zakharov V.N., Rogdestvenski Y.V., Diachenko Y.G., Morozov N.V., Stepchenkov D.Y. Speed-Independent Floating Point Coprocessor. Conference: 2015 East-West Design \& Test Symposium (EWDTS 2015), Batumi, Georgia, September 26-29, 2015, p. 111.
	
	\bibitem{Boulychev}
	Boulychev~D. , Koznov~D. , Terekhov~A.A. On project-specific languages and their application in reengineering. Software Maintenance and Reengineering, 2002. Proceedings. Sixth European Conference on Software Maintenance and Reengineering. 2002. P. 177.
	
	
	\bibitem{iarieli}
	Itai Arieli, Transfer Implementation in Congestion Games, Discussion Paper No. 9-14, October 2014, pp.17.
	
	\bibitem{tekin}
	Tekin~C., Liu~M., Southwell~R., Huang~J., Ahmad~S. Atomic Congestion Games on Graphs and Their Applications in Networking, IEEE Trans. on Networking, vol. 20, no. 5, pp. 1541-1552, October 2012. 
	
	\bibitem{Voevodin}
	Voevodin~V.V., Voevodin~V.V. Parallel computing, BHV, 2002, pp. 599. (in Russian).
	
	\bibitem{MalKol}
	Kolokoltsov~ V.N., Malafeyev~O.A. Understanding Game Theory. Introduction to the analysis of many agent systems of competition and cooperation, World Scientific, 2010, pp. 286.
	
	\bibitem{ershova}Ershova T.A., Malafeev O.A., Konfliktnye upravlenija v modeli vhozhdenija v rynok, Problemy mehaniki i upravlenija: Nelinejnye dinamicheskie sistemy. 2004. № 36. p. 19-27. 
	
	\bibitem{pahar}Malafeev O.A., Pahar O.V., Dinamicheskaja nestacionarnaja zadacha investirovanija proektov v uslovijah konkurencii, Problemy mehaniki i upravlenija: Nelinejnye dinamicheskie sistemy. 2009. № 41. p. 103-108. 
	
	\bibitem{murviov}Malafeev O.A., Murav'ev A.I., Matematicheskie modeli konfliktnyh situacij i ih razreshenie, Sankt-Peterburg, 2001. Tom 2 Matematicheskie osnovy modelirovanija processov konkurencii i konfliktov v social'no-jekonomicheskih sistemah, 294 p.
	
	\bibitem{alferov}Alferov G.V., Malafeyev O.A., Maltseva A.S., Programming the robot in tasks of inspection and interception, 2015 International Conference on Mechanics - Seventh Polyakhov's Reading 2015. p. 7106713. 
	
	\bibitem{sotnikova} Malafeev O.A., Sotnikova N.N. i dr. Linejnaja algebra s prilozhenijami k modelirovaniju korrupcionnyh sistem i processov, uchebnoe posobie / Stavropol', 2016, p.366.
	
	\bibitem{gricai}Malafeev O.A., Gricaj K.N., Konkurentnoe upravlenie v modeljah aukcionov, Problemy mehaniki i upravlenija: Nelinejnye dinamicheskie sistemy. 2004. № 36. p. 74-82. 
	
	\bibitem{gricai}Gricaj K.N., Malafeev O.A., Zadacha konkurentnogo upravlenija v modeli mnogoagentnogo vzaimodejstvija aukcionnogo tipa, Problemy mehaniki i upravlenija: Nelinejnye dinamicheskie sistemy. 2007. № 39. p. 36-45. 
	
	\bibitem{parf}Parfenov A.P., Malafeev O.A., Ravnovesnoe i kompromissnoe upravlenie v setevyh modeljah mnogoagentnogo vzaimodejstvija, Problemy mehaniki i upravlenija: Nelinejnye dinamicheskie sistemy. 2007. № 39. p. 154-167. 
	
	\bibitem{kefeli}Kefeli I.F., Malafeev O.A., Matematicheskie nachala global'noj geopolitiki, Sankt-Peterburg, 2013, 204 s.
	
	\bibitem{kolokoltsov} Kolokoltsov V.N., Malafeyev O.A., Understanding game theory: introduction to the analysis of many agent systems with competition and cooperation,  Understanding Game Theory: Introduction to the Analysis of Many Agent Systems with Competition and Cooperation 2010. p. 286. 
	
	
	


	
	
\end{thebibliography}
\end{document}